\documentclass[showkeys,eqnum,showpacs,prd,superscriptaddress]{revtex4}
\usepackage[dvips]{graphicx}
\usepackage{color}
\usepackage{amsmath}
\usepackage[dvips]{graphicx}
\begin{document}

\title{On Distance and Area}

\author{Jarmo M\"akel\"a} 

\email[Electronic address: ]{jarmo.makela@puv.fi}  
\affiliation{Vaasa University of Applied Sciences, Wolffintie 30, 65200 Vaasa, Finland}

\begin{abstract} 

We seek for an alternative to the metric tensor $g_{\mu\nu}$ as a fundamental geometrical object
in four-dimensional Riemannian manifolds. We suggest that the metric tensor $g_{\mu\nu}(P)$
at a given point $P$ of a manifold may be replaced by a four-dimensional geometrical 
simplex $\sigma^4(P)$ embedded to the tangent space $T_P$ of the point $P$. The number of 
two-faces, or triangles, of $\sigma^4(P)$ is the same as is the number of independent components
of $g_{\mu\nu}(P)$, and hence we may replace the components of $g_{\mu\nu}(P)$  by the two-face 
areas of $\sigma^4(P)$. In this sense the concept of distance may, in four-dimensional
Riemannian manifolds, be reduced to the concept of area. This result may find some applications
in the thermodynamical approaches to quantum gravity.
  
\end{abstract}

\pacs{04.20.Cv}
\keywords{metric tensor, four-simplex}

\maketitle

    Traditionally, the metric tensor $g_{\mu\nu}$ has been used as the object describing the
geometric properties of Riemannian manifolds. By means of the metric tensor we may calculate, 
among other things, lengths of curves, distances between points, and the change experienced by a
vector when it is parallel transported around a closed loop on a Riemannian manifold. Although 
the metric tensor is a natural choice for a fundamental geometrical object, however, it is by 
no means unique, and one may ask, whether the metric tensor could be replaced, in a natural 
manner, by some other objects carrying exactly the same geometrical information.

   In this paper we shall introduce an object with a potential of satisfying these 
requirements in four-dimensional Riemannian manifolds. Our idea is to replace the metric 
tensor $g_{\mu\nu}(P)$, which determines the inner product between the vectors of the tangent
space $T_P$ associated with a given point $P$ of the manifold, by a {\it four-dimensional 
geometrical simplex} $\sigma^4(P)$, which is embedded to the tangent space $T_P$. By 
definition, an $n$-dimensional geometrical simplex $\sigma^n$ is a convex hull of $(n+1)$
linearly independent points $v_0, v_1,..., v_n$ of a Eulidean space $\Re^m$ $(m\ge n)$, 
and it is denoted by $\sigma^n = v_0v_1...v_n$. \cite{yksi} The points $v_0, v_1,..., v_n$ are known
as the {\it vertices} of $\sigma^n$. Hence the simplex $\sigma^4(P)$ is a convex
hull of five linearly independent points $v_0(P), v_1(P),..., v_4(P) \in T_P$. In the tangent
space $T_P$ which, by definition, is a flat, Euclidean four-space we introduce a system of 
coordinates $\tilde{x}^\mu$ $(\mu = 0, 1, 2, 3)$ such that straight lines parallel to the
tangent vectors $\vec{b}_\mu(P)$ of the coordinate curves associated with the coordinates
$x^\mu$ of the points of the manifold at the point $P$ act as coordinate axes. In this 
system of coordinates an edge vector joining the vertices $v_a(P)$ and $v_b(P)$ of 
$\sigma^4(P)$ is of the form
\begin{equation}
\vec{l}_{ab}(P) = (\tilde{x}^\mu(b) - \tilde{x}^\mu(a))\vec{b}_\mu(P),
\end{equation}
where $\tilde{x}^\mu(a)$ and $\tilde{x}^\mu(b)$, respectively, are the coordinates of the 
vertices $v_a(P)$ and $v_b(P)$ $(a, b = 0, 1, 2, 3, 4)$, and we have used Einstein's sum rule.

   Consider now what happens, if we keep the coordinates $\tilde{x}^\mu(a)$ of the vertices of
$\sigma^4(P)$ as fixed and move the point $P$ around on the manifold. If the manifold is curved,
the tangent vectors $\vec{b}_\mu(P)$ will change when the point $P$ is moved and, as a 
consequence, the edge vectors of the simplex $\sigma^4(P)$ will also change. In other words, 
the properties of the four-simplex $\sigma^4(P)$ are different in different points of a curved 
manifold. The main idea of this paper is to relate the geometrical properties of the manifold to the
geometrical properties of $\sigma^4(P)$.

   It is interesting that the number of the two-faces, or triangles, of the simplex $\sigma^4(P)$
is 
\begin{equation}
\left(\begin{array}{ccc}
5\\
3\end{array}\right) = 10,
\end{equation}
which is exactly the same as is the number of independent components of the metric tensor
$g_{\mu\nu}(P)$ at the point $P$. In other words, there is a one-to-one correspondence between
the two-faces of $\sigma^4(P)$ and the components of $g_{\mu\nu}(P)$. The natural geometrical 
quantities to replace the components of $g_{\mu\nu}(P)$ are therefore the areas of the two-faces
of $\sigma^4(P)$. When the components of $g_{\mu\nu}(P)$ are changed, the areas of the
two-faces of $\sigma^4(P)$ will also change and there is a one-to-one correspondence between 
the changes of the components of $g_{\mu\nu}(P)$ and the changes of the areas of the two-faces
of $\sigma^4(P)$. This means that exatly the same geometrical information is carried by the 
components of $g_{\mu\nu}(P)$ and the areas of the two-faces of $\sigma^4(P)$.

   In general, the relationship between the components of $g_{\mu\nu}(P)$ and the areas of the
two-faces of $\sigma^4(P)$ is pretty complicated. However, between the infinitesimal variations
of the two-face areas of $\sigma^4(P)$ and those of the components of $g_{\mu\nu}(P)$ there is 
a simple linear relationship. More precisely, if we arrange the infinitesimal variations $\delta
g_{\mu\nu}(P)$ to a column matrix $\delta g(P)$ with 10 elements, and those of the areas of the 
two-faces of $\sigma^4(P)$ to a column matrix $\delta A(P)$, we have:
\begin{equation}
\delta A(P) = M(P)\,\delta g(P)
\end{equation}
where $M(P)$ is an approriate $10\times 10$ matrix defined at the point $P$. Eq.(3) tells 
in which way the variations of the two-face areas may be obtained from the variations of the 
components of $g_{\mu\nu}(P)$. Conversely, the variations of the components of $g_{\mu\nu}(P)$
may be obtained by means of the variations of the two-face areas of $\sigma^4(P)$:
\begin{equation}
\delta g(P) = N(P)\,\delta A(P),
\end{equation}
where the $10\times 10$ matrix $N(P)$ is the inverse of the matrix $M(P)$. Eq.(4) implies that 
the partial derivatives of the components of the metric tensor may be expressed in terms of the 
partial derivatives of the two-face areas of $\sigma^4(P)$:
\begin{equation}
\frac{\partial g(P)}{\partial x^\mu} = N(P)\,\frac{\partial A(P)}{\partial x^\mu}
\end{equation}
for all $\mu = 0, 1, 2, 3$. If we pick up an orthonormal system of coordinates at the point $P$,
the first partial derivatives of the metric tensor will all vanish at $P$, and its second partial
derivatives may all be expressed in terms of the second partial derivatives of the two-face areas:
\begin{equation}
\frac{\partial^2 g(P)}{\partial x^\mu\partial x^\nu} 
= N(P)\,\frac{\partial^2A(P)}{\partial x^\mu\partial x^\nu}.
\end{equation}
Since the Riemann tensor and the related objects such as the Ricci and the Einstein tensors are
all, in orthonormal geodesic coordinates, functions of the second partial derivatives of the 
components of the metric tensor only, we find that all these objects may be expressed in terms
of the second partial derivatives of the two-face areas of $\sigma^4(P)$. In an arbitrary system 
of coordinates the components of these objects may be obtained from their components in  
orthonormal geodesic coordinates by means of a simple cordinate trasformation. In other words,
we have shown that in an arbitrary point $P$ of a Riemannian manifold in an arbitrary system of 
coordinates the components of the Riemann tensor, and thus all geometrical properties of the 
manifold at that point, may ultimately be reduced to the two-face areas of $\sigma^4(P)$. Hence 
we may indeed replace the metric tensor $g_{\mu\nu}$ as a fundamental geometric object of 
four-dimensional Riemannian manifolds by a specific four-simplex $\sigma^4$.

  It only remains to find the $10 \times 10$ matrices $M(P)$ and $N(P)$. In a proper Riemannian
manifold with a positive definite metric tensor the area of a two-simplex with vertices $v_a(P)$, 
$v_b(P)$ and $v_c(P)$ is
\begin{equation}
A_{abc}(P) = \frac{1}{4}\sqrt{4s_{ab}(P)s_{ac}(P) - (s_{ab}(P) + s_{ac}(P) - s_{bc}(P))^2},
\end{equation}
where 
\begin{equation}
s_{ab}(P) := g_{\mu\nu}(P)(\tilde{x}^\mu(b) - \tilde{x}^\mu(a))(\tilde{x}^\nu(b) - \tilde{x}^\nu(a))
\end{equation}
is the squared length of the edge vector joining the vertices $v_a(P)$ and $v_b(P)$. Hence we 
find that
\begin{equation}
\delta A_{abc} = M^{\mu\nu}_{abc}(P)\,\delta g_{\mu\nu}(P),
\end{equation}
where
\begin{equation}
M^{\mu\nu}_{abc}(P) := \frac{1}{16A_{abc}(P)}[\Delta_{abc}(P)(ab)^{\mu\nu} 
                                             + \Delta_{cab}(P)(ca)^{\mu\nu}
                                             + \Delta_{bca}(P)(bc)^{\mu\nu}].
\end{equation}
In Eq.(10) we have denoted:
\begin{subequations}
\begin{eqnarray}
\Delta_{abc}(P) &:=& s_{ac}(P) + s_{bc}(P) - s_{ab}(P),\\
(ab)^{\mu\nu} &:=& (\tilde{x}^\mu(b) - \tilde{x}^\mu(a))(\tilde{x}^\nu(b) - \tilde{x}^\nu(a)).
\end{eqnarray}
\end{subequations}
The quantities $M^{\mu\nu}_{abc}(P)$ are the elements of the matrix $M(P)$. The indices $\mu\nu$ 
determine collectively the column and the indices $abc$ the row of the element 
$M^{\mu\nu}_{abc}(P)$. In a pseudo-Riemannian manifold with a signature $(-,+,+,+)$ in the metric
the elements of $M^{\mu\nu}_{abc}(P)$ are otherwise the same as in Eq.(10), except that for 
time-like two-faces the right hand side of Eq.(10) is equipped with a minus sign and the expression 
inside of the square root in Eq.(7) is replaced by its modulus. The elements of the inverse
$N(P)$ of the matrix $M(P)$ are of the form $N^{abc}_{\mu\nu}(P)$, and they have the property:
\begin{equation}
\delta g_{\mu\nu}(P) = N^{abc}_{\mu\nu}(P)\,\delta A_{abc}(P).
\end {equation}
In contrast to the elements of the matrix $M(P)$, the indices $abc$ determine collectively the
column and the indices $\mu\nu$ the row of the element $N^{abc}_{\mu\nu}(P)$ of $N(P)$. 

   When calculating the elements of the matrices $M(P)$ and $N(P)$ we have lots of choice, 
because those elements depend both on the four-simplex $\sigma^4(P)$ chosen at the point $P$,
and on the system of coordinates. In the Appendix we have constructed the matrices $M(P)$ and 
$N(P)$ explicitly in the special case, where the lengths of all edges of $\sigma^4(P)$ are 
equals, and the system of the coordinates $x^\mu(P)$ at the point $P$ of the manifold 
has been chosen such that in the corresponding coordinates $\tilde{x}^\mu$ induced in the tangent space $T_P$ the
vertex $v_4(P)$ lies at the origin,  and the coordinates of the other vertices $v_a(P)$ are 
$\tilde{x}^\mu(a) = \delta^\mu_a$ for all $a, \mu = 0, 1, 2, 3$. If we change the "old" coordinates
$x^\mu$ to the "new" coordinates $x'^\mu$, the elements of the matrices $M(P)$ and $N(P)$ will
also change such that
\begin{subequations}
\begin{eqnarray}
M'^{\mu\nu}_{abc}(P) &=& \frac{\partial x'^\mu}{\partial x^\alpha}\frac{\partial 
x'^\nu}{\partial x^\beta}M^{\alpha\beta}_{abc}(P),\\
N'^{abc}_{\mu\nu}(P) &=& \frac{\partial x^\alpha}{\partial x'^\mu}
\frac{\partial x^\beta}{\partial x'^\nu}N^{abc}_{\alpha\beta}(P).
\end{eqnarray}
\end{subequations}
So it is possible to obtain the elements of the matrices $M(P)$ and $N(P)$ in any system
of coordinates, provided that we know those elements in just one system of coordinates.

   In this paper we have suggested that the metric tensor $g_{\mu\nu}(P)$ determining the inner
product between the vectors of the tangent space $T_P$ associated with a given point $P$ of 
a Riemannian manifold may be replaced, in four dimensions, by a specific four-simplex 
$\sigma^4(P) \subset T_P$. In the tangent space $T_P$ we define a system of coordinates, where
straight lines parallel to the tangent vectors of the coordinate curves of the manifold act as 
coordinate axes. In this system of coordinates we keep the coordinates of the vertices of 
$\sigma^4(P)$ as fixed. When the point $P$ is moved around on a curved manifold, the tangent 
vectors of the coordinate curves, and hence the edges of the four-simplex $\sigma^4(P)$, will
change. Four-dimensional simplices have a specific property that the number of their two-faces,
or triangles, is the same as is the number of independent components of the metric tensor.
Hence there is a one-to-one correspondence between the changes of the two-face areas of 
$\sigma^4(P)$ and the changes of the independent components of the metric tensor $g_{\mu\nu}(P)$,
when the point $P$
is moved around on the manifold. In other words, the changes of the components of the metric
tensor $g_{\mu\nu}(P)$ may be expressed in terms of the changes of the two-face areas of 
$\sigma^4(P)$, and vice versa. Because of that we may replace the components of the metric
tensor, which determines the distances between nearby points, by the two-face areas of a 
specific four-simplex as the fundamental geometrical variables
in four-dimensional Riemannian manifolds. In this sense one may say that in four-dimensional
Riemannian manifolds the concept of distance may be reduced to the concept of area.

  The explicit relationship between the components of the metric tensor and the two-face 
areas of our four-simplex is, in general, pretty complicated, and therefore it is unlikely that 
the use of two-face areas of a four-simplex would offer essential benefits over the use of the
components of the metric tensor in the traditional applications, such as classical general 
relativity, of the general theory of Riemanian manifolds. Nevertheless, it is quite interesting
that the concept of metric tensor, which determines the distance between nearby points, may be 
replaced by the concept of area in the sense described above. The potential importance of this
result lies in the fact that in many approaches to quantum gravity the concept of area, instead 
of the concepts of metric and distance, takes a central role. In loop quantum gravity, 
for example, spacetime is assumed to consist of Planck-size loops equipped with a certain area 
spectrum. \cite{kaksi} During some recent times attempts to consider general relativity as an essentially
thermodynamical theory of spacetime and its constituents have gained increasing popularity. 
\cite{kolme, nelja, viisi, kuusi}
In those considerations the concept of entropy holds the central stage. Because the entropy
of a black hole is proportional to its event horizon area, one may expect the concept of area
to play a fundamental role in any thermodynamical approach to quantum gravity. For instance,
the results gained from the thermodynamical approaches to quantum gravity so far suggest that
two-dimensional surfaces of spacetime might consist of Planck-size constituents, each of them 
occupying an area which is about one Planck length squared. \cite{viisi, kuusi, kahdeksan} It is possible that by means of 
the results of this paper one may obtain a relationship between the geometric and the causal
poperties of spacetime, and the statistical distributions of those constituents in their 
quantum states. 

\appendix*

\section{The Matrices M(P) and N(P)}

  In this appendix we shall obtain explicit expressions for the matrices $M(P)$ and $N(P)$
in the special case, where the edge lengths of the geometric four-simplex $\sigma^4(P)$ are
all equals and the coordinates $\tilde{x}^\mu(a)$ of the vertices $v_a(P)$ of the simplex
in the tangent space $T_P$ have been chosen in such a way that
\begin{equation}
\tilde{x}^\mu(a) := \delta^\mu_a,
\end{equation}
for all $a = 0, 1, 2, 3$, and 
\begin{equation}
\tilde{x}^\mu(4) = 0.
\end{equation}
In other words, the vertex $v_4(P)$ lies at the origin of our system of coordinates, and
the coordinates of the vertices $v_0(P), v_1(P), v_2(P)$ and $v_3(P)$, respectively, are
$(1,0,0,0)$, $(0,1,0,0)$, $(0,0,1,0)$ and $(0,0,0,1)$. Denoting the common length of the
edges of $\sigma^4(P)$  by $L$ we find:
\begin{subequations}
\begin{eqnarray}
A_{abc}(P) &=& \frac{\sqrt{3}}{4}L^2,\\
\Delta_{abc}(P) &=& L^2
\end{eqnarray}
\end{subequations}
for all $a, b, c = 0, 1, 2, 3, 4$. Hence it follows fom Eq.(10) that the elements of the matrix
$M(P)$ are:
\begin{equation}
M^{\mu\nu}_{abc}(P) = \frac{\sqrt{3}}{12}[(ab)^{\mu\nu} + (ca)^{\mu\nu} + (bc)^{\mu\nu}].
\end{equation}

   Using Eq.(11b) we observe that the symbols $(ab)^{\mu\nu}$ have the symmetry properties:
\begin{subequations}
\begin{eqnarray}
(ab)^{\mu\nu} &=& (ab)^{\nu\mu},\\
(ab)^{\mu\nu} &=& (ba)^{\mu\nu},
\end{eqnarray}
\end{subequations}
and therefore the only independent, non-zero components of $(ab)^{\mu\nu}$, in our system of coordinates,
are:
\begin{subequations}
\begin{eqnarray}
(ab)^{aa} &=& 1, \,\,\, (a = 0, 1, 2, 3)\\
(ab)^{ab} &=& -1. \,\,\, (a, b = 0, 1, 2, 3)
\end{eqnarray}
\end{subequations}
So we find that if we define the column matrices $\delta A(P)$ and $\delta g(P)$ such that
\begin{equation}
\delta A(P) := \left(\begin{array}{ccc}
\delta A_{012}(P)\\
\delta A_{013}(P)\\
\delta A_{014}(P)\\
\delta A_{023}(P)\\
\delta A_{024}(P)\\
\delta A_{034}(P)\\
\delta A_{123}(P)\\
\delta A_{124}(P)\\
\delta A_{134}(P)\\
\delta A_{234}(P)
\end{array}\right)
\quad\text{and}\quad
\delta g(P) := \left(\begin{array}{ccc}
\delta g_{00}(P)\\
\delta g_{01}(P)\\
\delta g_{02}(P)\\
\delta g_{03}(P)\\
\delta g_{11}(P)\\
\delta g_{12}(P)\\
\delta g_{13}(P)\\
\delta g_{22}(P)\\
\delta g_{23}(P)\\
\delta g_{33}(P)
\end{array}\right),
\end{equation}
the matrix $M(P)$ becomes to:
\begin{equation}
M(P) = \frac{\sqrt{3}}{12}\left(\begin{array}{cccccccccc}
2&-1&-1&0&2&-1&0&2&0&0\\
2&-1&0&-1&2&0&-1&0&0&2\\
2&-1&0&0&2&0&0&0&0&0\\
2&0&-1&-1&0&0&0&2&-1&2\\
2&0&-1&0&0&0&0&2&0&0\\
2&0&0&-1&0&0&0&0&0&2\\
0&0&0&0&2&-1&-1&2&-1&2\\
0&0&0&0&2&-1&0&2&0&0\\
0&0&0&0&2&0&-1&0&0&2\\
0&0&0&0&0&0&0&2&-1&2
\end{array}\right).
\end{equation}
The matrix $N(P)$ is the inverse of $M(P)$:
\begin{equation}
N(P) = \frac{2\sqrt{3}}{3}\left(\begin{array}{cccccccccc}
-1&-1&2&-1&2&2&2&-1&-1&-1\\
-4&-4&2&2&2&2&2&2&2&4\\
-4&2&2&-4&2&2&2&2&-4&2\\
2&-4&2&-4&2&2&2&-4&2&2\\
-1&-1&2&2&-1&-1&-1&2&2&-1\\
-4&2&2&2&2&-4&-4&2&2&2\\
2&-4&2&2&-4&2&-4&2&2&2\\
-1&2&-1&-1&2&-1&-1&2&-1&2\\
2&2&-4&-4&2&2&-4&2&2&2\\
2&-1&-1&-1&-1&2&-1&-1&2&2\\
\end{array}\right).
\end{equation}
Using the matrix $M(P)$ we may express the elements of $\delta A(P)$ in terms of the 
elements of $\delta g(P)$, whereas by means of the matrix $N(P)$ we may express the 
elements of $\delta g(P)$ in terms of the elements of $\delta A(P)$. So there is 
an invertible one-to-one relationship between the variations $\delta g_{\mu\nu}(P)$ and
$\delta  A_{abc}(P)$ of the components of the metric tensor and the areas of the 
two-faces of $\sigma^4(P)$. The elements of the matrices $M(P)$ and $N(P)$ in any 
system of coordinates may be obtained from the elements of $M(P)$ and $N(P)$ in Eqs.(A8)
and (A9) by means of a simple coordinate transformation as in Eqs. (13a) and (13b). 
Although we have assumed in this Appendix that our manifold is a proper Riemannian 
manifold with a positive definite metric, a similar calculation may be performed in 
pseudo-Riemannian manifolds as well.


\begin{thebibliography} {20}

\bibitem{yksi} M. K. Agoston, {\it Algebraic Topology: A First Course} 
(Marcel Dekker, New York, 1976).

\bibitem{kaksi} For an introduction to loop quantum gravity see, for example, P. Dona and 
S. Speziale, arXiv:1007.0402

\bibitem{kolme} T. Jacobson, Phys. Rev. Lett. {\bf 75} (1995) 1260.

\bibitem{nelja} J. M\"akel\"a and A. Peltola, Int. J. Mod. Phys. {\bf D18} (2009) 669.

\bibitem{viisi} E. Verlinde, arXiv:1001.0785

\bibitem{kuusi} J. M\"akel\"a, arXiv:1001.3808

\bibitem{seitseman} For a general review and bibliography see, for instance, 
T. Padmanabhan, Rep. Prog. Phys. {\bf 73} (2010) 046901. Also see,
T. Padmanabhan, Mod. Phys. Lett. {\bf A25} (2010) 1129, and  Phys. Rev. {\bf D81} (2010) 124040.

\bibitem{kahdeksan} J. M\"akel\"a, arXiv:0810.4910 

\end{thebibliography}
\end{document}